\documentclass{webofc}
\usepackage[varg]{txfonts}   
\newcommand\beq{\begin{equation}}
\newcommand\eeq{\end{equation}}
\newcommand\beqa{\begin{eqnarray}}
\newcommand\eeqa{\end{eqnarray}}

\newcommand\bgamma{\mbox{\boldmath$\gamma$}}

\newcommand\bq{{\bf q}}

\newcommand\bx{{\bf x}}
\newcommand\br{{\bf r}}
\newcommand\bk{{\bf k}}

\newcommand\balpha{\mbox{\boldmath$\alpha$}}
\newcommand\bP{{\bf P}}

\woctitle{International Conference on New Frontiers in Physics 2013}
\begin{document}
\title{Inhomogeneous chiral phase in the magnetic field}
%
%

\author{T.Tatsumi\thanks{\email{tatsumi@ruby.scphys.kyoto-u.ac.jp}} \and
        K. Nishiyama \and
        S. Karasawa
             }

\institute{Department of Physics, Kyoto University, Kyoto 606-8502, Japan 
          }

\abstract{%
Inhomogeneous chiral phase is discussed in the presence of the magnetic field. 
A topological aspect is pointed out for the complex order parameter, 
in relation to the spectral asymmetry of the Dirac operator. 
It induces an anomalous baryon number and extremely extends the region of 
the inhomogeneous chiral phase in the QCD phase diagram. 
It is also shown that the novel tricritical point appears at zero chemical potential,
which should be examined by the lattice QCD simulation.
}
\maketitle
\section{Introduction}
\label{Intro}

The QCD phase diagram has been studied by theoretical and experimental approaches \cite{fuk}.
Recently possibility of the inhomogeneous chiral phases has attracted much interest in the QCD phase diagram. They are specified by the spatially modulating quark condensates: the order parameter of the chiral transition is now generalized to be complex function,
$\langle{\bar\psi}\psi\rangle+i\langle{\bar\psi}i\gamma_5\tau_3\psi\rangle
\equiv \Delta(\br){\rm exp}\left( i\theta(\br)\right)$, within $SU(2)_L\times SU(2)_R$ chiral symmetry.
The studies using the NJL model have shown the appearance of the inhomogeneous phases in the vicinity of the chiral transition. Two typical forms of the condensates have been well studied: one is dual chiral density wave (DCDW) specified by $\Delta={\rm const.}, \theta=\bq\cdot\br$ \cite{nak} and the real kink crystal (RKC) 
specified by $\Delta(\br), \theta={\rm const.}$ \cite{bas,nic}. 
Nowadays it is known that we can generally consider the inhomogeneous phases with one dimensional order by embedding the Hartree-Fock solutions obtained 
in 1+1 dimensions \cite{bas} in 1+3 dimensions \cite{nic}.

The spatially dependent order parameter is not special and rather common in condensed matter physics: examples are the FFLO state in superconductivity \cite{fflo} and spin density wave \cite{ove} or texture structure in magnetism \cite{con}. Similar subject has been also discussed in the context of color superconductivity 
\cite{alf}.

Here we discuss the properties of the 
inhomogeneous chiral phases in the presence of the external magnetic field. The magnetic field is popular in hadron physics and is found in 
various situations as in pulsars. The generation of huge magnetic field of $O(10^{17}{\rm G})$ has been recently advocated during the relativistic heavy-ion collisions. The magnetic field may affect the phase transition or induce new phenomena, and there have been many theoretical works  \cite{lec,bal}. 
Thus our purpose is to explore the 
inhomogeneous chiral phases in the space of  
temperature ($T$)-chemical potential ($\mu$)-magnetic field ($H$). 
    
When the magnetic field is present, the energy spectrum is quantized as the Landau levels. 
The spacing of the Landau levels becomes large as $H$ increases, 
so that the dimensional reduction is realized in the large $H$ limit: 
there is left only one dimensional degree of freedom along the magnetic field. 
More interestingly, the lowest Landau level exhibits spectral asymmetry for some kinds of the inhomogeneous chiral phases and induces 
anomalous baryon number through the Atiyah-Patodi-Singer $\eta$ invariant \cite{nie}, which is a topological object and homotopy invariant.
 We emphasize such topological aspect of the inhomogeneous chiral phases.
    
\section{Spectral asymmetry}
\label{sec-1}

For the phase transitions of the homogeneous order parameters, it has been well-known that the spontaneous symmetry breaking (SSB) is 
enhanced by the magnetic field \cite{kle,sug}, and it is sometimes called {\it magnetic catalysis} \cite{gus}. Here we consider the inhomogeneous phases in the presence of the magnetic field.
We can construct the inhomogeneous chiral phase by embedding the general solutions given in 1+1 dimensions $M(z)$ in 1+3 dimensions \cite{nic}.
Then the Dirac operator is given by ${\cal H}_D=\balpha\cdot\bP+\gamma^0 \left[\frac{1+\gamma_5\tau_3}{2}M(z)+\frac{1-\gamma_5\tau_3}{2}M^*(z)\right]$ 
with $M(z)=-2G(\langle{\bar\psi}\psi\rangle+i\langle{\bar\psi}i\gamma_5\tau_3\psi\rangle)$ in the presence of the magnetic field, where $G$ is the coupling constant of the NJL model, $\balpha=\gamma^0\bgamma$ and $\bP=-i\nabla+Q{\bf A}$ with the em charge $Q={\rm diag} (e_u,e_d)$. 
We apply the magnetic field along the $z-$ axis, which should be the most favorite case \cite{fro}.
Using the Landau gauge, ${\bf A}=(0,Hx,0)$, ${\cal H}_D$ then can be represented as, for each flavor,  
\begin{equation}
{\cal H}_D=\left(
\begin{array}{cccc}
-i\partial_z &M(z)         &0            &\sqrt{2|e_fH|n}\\
M^*(z)       &i\partial_z  &-\sqrt{2|e_fH|n} &0         \\
0            &-\sqrt{2|e_fH|n} &-i\partial_z &M^*(z)        \\ 
\sqrt{2|e_fH|n}  &0            &M(z)         &i\partial_z
\end{array}
\right),
\label{dirac}
\end{equation}
on the basis of the product of the plane wave ${\rm exp}(iky)$ and the Hermite function $u_n(\xi)$ with $\xi=\sqrt{|e_fH|}x+k/\sqrt{|e_fH|}$ \cite{sok}, where $n$ denotes the Landau levels. 
Thus, we can discuss general configurations $M(z)$ in the presence of the magnetic field. 

\subsection{Dual chiral density wave}

\begin{figure}
\centering
\sidecaption
\includegraphics[width=5cm,clip]{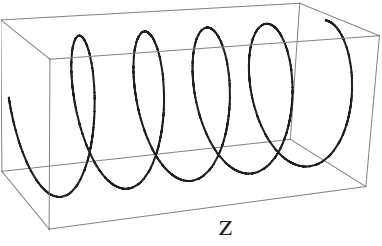}
\caption{Sketch of DCDW, which develops along the horizontal axis $z$. The vertical plane is spanned by two axes of $\langle {\bar \psi}\psi\rangle$ and 
$\langle {\bar \psi}i\gamma_5\tau_3\psi\rangle$.}
\label{fig-1}       
\end{figure}

In the following, we focus on DCDW \cite{nak}, which is specified by the uniform amplitude $\Delta$ and the chiral angle $\theta(z)=qz$ (see Fig.~1); 
\beqa
\langle{\bar \psi}\psi\rangle&=&\Delta\cos(qz),\nonumber\\
\langle{\bar \psi}i\gamma_5\tau_3\psi\rangle&=&\Delta\sin(qz),
\label{dcdw}
\eeqa
which utilizes only $U(1)$ subgroup in the isospin $SU(2)$ to ensure the state to be charge eigenstate. The state vector for DCDW can be given by operating the 
 {\it local} chiral transformation on the normal state,
\begin{eqnarray}
|{\rm DCDW}\rangle=&=&{\rm exp}\left(i\int\theta(\br) A_3^0(\br)d^3r\right)\left.\right|{\rm normal}\rangle\nonumber\\
&\equiv &U_{\rm DCDW}(\theta(\br))\left.\right|{\rm normal}\rangle,
\end{eqnarray}
with the chiral angle $\theta(\br)=qz$, where $A_i^\mu$ denotes the axial-vector current with $i$-th isospin component. 
Actually one can easily check (\ref{dcdw}) with $\Delta=\langle {\rm normal}\left|{\bar \psi}\psi\right|{\rm normal}\rangle$. 
We choose this among general configurations, since we can expect 
DCDW becomes the most favorite configuration in the strong magnetic field, once the dimensional reduction efficiently works \cite{bas}. 


The Lifshitz point for DCDW meets the tricritical point (TCP) of the usual chiral transition in the chiral limit, and its emergence may be 
qualitatively understood in terms of the nesting effect of the Fermi surface \cite{nak}. In the recent work, theoretical framework has been naturally 
extended to take into account the current quark mass $m_c$ \cite{kar}. Using a variational method, 
the function form of the chiral angle is deformed to satisfy the sine-Gordon equation. 
Taking the NJL model and using the proper-time regularization we can depict the phase diagram in the $T-\mu$ plane, presented in Fig.~2 \cite{kar}.

\begin{figure}
\centering
\includegraphics[width=6cm,clip]{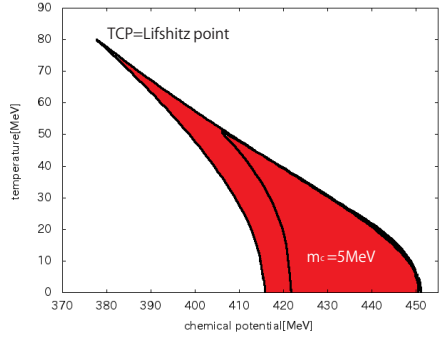}
\caption{Phase diagram in the $\mu-T$ plane. The larger colored domain denotes the DCDW phase in the chiral limit, $m_c=0$, while the smaller domain corresponds to the case of $m_c=5$MeV \cite{kar}.}
\label{fig-2}       
\end{figure}

\subsection{DCDW in the magnetic field}

For DCDW  it is easy to diagonalize the Hamiltonian (\ref{dirac}), and  
the energy spectrum then can be expressed as 
\beqa
E_{n,p,\zeta=\pm 1,\epsilon=\pm 1}&=&\epsilon\sqrt{\left(\zeta\sqrt{m^2+p^2}+q/2\right)^2+2|e_fH|n}, n=1,2,..., \nonumber\\ 
E_{n=0,p,\epsilon=\pm 1}&=&\epsilon\sqrt{m^2+p^2}+q/2,
\label{spec}
\eeqa 
for each flavor \cite{fro}, where $m$ denotes the dynamical mass defined by $m=-2G\Delta$ within the NJL model. 
For the lowest Landau level (LLL), $n=0$, ${\cal H}_D$ is reduced to $2\times 2$ matrix,
\beq
H_{LLL}(M(z))=\left(
\begin{array}{cc}
-i\partial_z &M(z)    \\
M^*(z)       &i\partial_z 
\end{array}
\right)
\eeq
from the property of the Hermite function $u_n(\xi)$, $u_{-1}=0$. The spectrum of LLL then exhibits spectral asymmetry, which induces the anomalous baryon number \cite{nie}. 
The anomalous baryon number has a topological origin; when the quark number operator is defined in the symmetric form, ${\hat N}=1/2\int d^3x[\psi^\dagger(\bx),\psi(\bx)]$, it is given by the Atiyah-Patodi-Singer $\eta$ invariant $\eta_H$, $N_{\rm anom}=-1/2\eta_H$. $\eta_H$ is defined by 
\beq
\eta_H=\lim_{s\rightarrow 0}\sum_\lambda |E_\lambda|^{-s}{\rm sign} (E_\lambda),
\eeq
where $E_\lambda$ generally denotes the eigenvalue of the Dirac operator, including flavor and color degrees of freedom. We can see that the order parameter $M$ must be complex to give a finite value of $\eta_H$; the phase degree of freedom (chiral angle) $\theta$ is essential in other words. It can be easily extended to the thermodynamic limit \cite{nie}, and quark number is given by 
\beq
N=-\frac{1}{2}\eta_H+\sum_\lambda{\rm sign}(E_\lambda)\left[\frac{\theta(E_\lambda)}{e^{\beta(E_\lambda-\mu)}+1}+
\frac{\theta(-E_\lambda)}{e^{-\beta(E_\lambda-\mu)}+1}\right],
\eeq
where the first term is the quark number of the Dirac sea, and the second term the one of the Fermi sea.
We then see that thermodynamic potential $\Omega$ should include the anomalous term as well through the thermodynamic relation, $N=-\partial\Omega/\partial\mu$.
It is to be noted that the anomalous baryon number is well known in the context of the chiral bag model: Goldstone and Jaffe showed the baryon number of the chiral bag is exactly one, once the spectral asymmetry is taken into account inside the bag \cite{gol,mul}.

We can directly evaluate the $\eta$ invariant, assuming $q/2<m$, which may be justified at least in the low density region $\mu<m$.
Using the Mellin transform,
\beq
\left|\lambda\right|^{-s}=\frac{1}{\Gamma(s)}\int_0^\infty d\omega\omega^{s-1}{\rm exp}\left(-|\lambda|\omega\right),
\eeq
\beqa
\eta_{\rm H}^f(s)&=&N_c\frac{2L}{\Gamma(s)}\frac{|e_fH|}{2\pi}\int\frac{dp}{2\pi}\int_0^\infty d\omega\omega^{s-1}\sum_{\epsilon}{\rm sign}(\lambda_{L,\epsilon}){\rm exp}\left(-\left|\lambda_{L,\epsilon}\right|\omega\right)\nonumber\\
&=&\sqrt{\pi q}\frac{i}{\Gamma(s)}\frac{|e_f H|}{4}m^{-s+1/2}{\rm F}_{1,1/2,s-1/2},
\eeqa
with 
\beqa
{\rm F}_{1,1/2,s-1/2}&=&i\frac{2^{s-1/2}(q/2m)^{1/2}}{\pi^{3/2}/2}\frac{s}{2}\Gamma\left(\frac{s}{2}\right)^2 
{_2{\rm F}_1}\left(1+\frac{s}{2},\frac{s}{2},\frac{3}{2}; \left(\frac{q}{2m}\right)^2\right).
\eeqa
Finally we have 
\beq
\eta_{\rm H}^f/(2L)=\lim_{s\rightarrow 0+}\eta_{\rm H}^f(s)/(2L)=-N_c\frac{|e_fH|q}{2\pi^2}.
\eeq
Note that it is independent of dynamical mass $m$. Anomalous baryon-number density is then 
\beq
\rho_{\rm anom}=\sum_f\frac{|e_f H|}{4\pi^2}q.
\eeq 
As we shall see in \ref{sec-2}, the wave vector $q$ should be zero at $\mu=0$ 
and then increases as $\mu$ in the magnetic field, 
which means that baryon-number density is always finite for any value of $\mu$. 
Accordingly DCDW may develop for  
any chemical potential in the presence of the magnetic field. Actually the results in ref.\cite{fro} 
look to support our conclusion.

\subsection{Comments on the chiral spiral in 1+1 dimensions}

Here we consider the chiral spiral in 1+1 dimensions in the absence of the magnetic field \cite{bas}.
The spectrum is then the same as the one of LLL (\ref{spec}), so that it exhibits spectral asymmetry for $q\neq 0$. 
We consider only one flavor for simplicity.
Topological baryon-number density then can be written as
\beq
\rho_{\rm anom}=\frac{q}{2\pi},
\label{ano}
\eeq  
for $m>q/2$, which is satisfied for small chemical potential $\mu$ \cite{bas}.  Thus 
the baryon-number density is non-vanishing for $q\neq 0$, irrespective of chemical potential in this region.
Note here that the relation $q=2\mu$ holds as the optimal value of $q$ for given $\mu$ \cite{bas}.
This sounds a little bit curious, since the dynamical mass $m$ is larger than $\mu$ there \cite{bar}.
However, the anomalous baryon-number density produces the Fermi sphere with the Fermi momentum $k_F$ defined through the relation, 
$\rho=k_F/\pi$. Using Eq.~(\ref{ano}), we can see $k_F=q/2$. 
This relation between $q$ and $k_F$ is exactly known as the nesting relation \cite{gru}: 
the nesting effect of thge Fermi surface is complete in 1+1 dimensions. 
Thus we can say that the chiral spiral appears as a result of the nesting effect and develops over all $\mu$ in 1+1 dimensions.
Moreover, one may expect the chiral spiral should be the most favorate configuration due to the nesting effect, compared with other ones.
Actually it has been numerically shown that the chiral spiral develops over any $\mu$ below the critical temperature 
and is the most favorite configuration in 1+1 dimensions \cite{bas}.
Finally, it should be interesting to observe the relation, $\mu=k_F$, which resembles 
the one for massless particles \cite{sch}.

\section{Novel tricritical point}
\label{sec-2}

One of the consequences of the spectral asymmetry is the appearance of the novel critical point in the $T-H$ plane.
Consider the generalized Ginzburg-Landau expansion in the vicinity of the phase transition. Thermodynamic potential can 
be expanded in therms of the order parameter and its derivatives \cite{thi},
\beqa
&&\Delta\Omega(M)
=\frac{\alpha_2}{2}\left|M\right|^2+\frac{\alpha_3}{3}{\rm Im}\left(MM'^*\right)
+\frac{\alpha_4}{4}\left(\left|M\right|^4+\left|M'\right|^2\right)\nonumber\\
&+&\frac{\alpha_5}{5}{\rm Im}\left(\left(M''-3|M|^2M\right)M'^*\right)
+\frac{\alpha_6}{6}\left(|M|^6+3|M|^2\left|M'\right|^2+2|M|^2\left|M^2\right|'+\frac{1}{2}\left|M''\right|^2\right)+...
\eeqa
This expression is obtained by using the NJL model, but its form should be model independent up to the third order term.
The coefficients $\alpha_l$ are functions of temperature, chemical potential and magnetic field; using Eq.(\ref{dirac}) we can find 
\beq
\alpha_l=(-1)^{l/2}8N_cT\sum_f\sum_m\int_{\rm reg}\frac{d^3k}{(2\pi)^3}{\rm tr}\left[\tilde{S}_A^f(k)\right]^l,
\eeq
where $\tilde{S}_A^f(k)$ is the quark Green function 
in the presence of the magnetic field, and can be further decomposed over the Landau levels,
\beq
{\tilde S}_A^f(k)=i{\rm exp}\left(-\frac{\bk_\perp^2}{|e_fH|}\right)\sum_{n=0}^{\infty}(-1)^n\frac{D_n(e_fH,k)}{(\omega_m+i\mu)^2-k_3^2-2|e_fH|n},
\eeq
with the Matsubara frequency, $\omega_m=(2m+1)\pi T$.
The numerator $D_l$ is a somewhat complicated function \cite{cho}. 
We can see that the odd-index terms obviously vanish for the real order parameter. The Lifshitz point is then given by the condition such that 
the leading even-index terms vanish, $\alpha_2=\alpha_4=0$, which is identical to the condition of the tricritical point 
in the usual case within the NJL model\cite{nic}.
On the other hand, the odd-index terms survive in the presence of the magnetic field due to the spectral asymmetry of LLL.
Accordingly we shall see the novel tricritical point appears, which is defined as the point with the condition,
\beq
\alpha_2=\alpha_3=0.
\eeq 
This condition may remind one of the chiral spiral in 1+1 dimensions, where the tricritical point resides on the $\mu=0$ line.
Actually $\alpha_3$ can be written as 
\beq
\alpha_3(H,T,\mu)=\frac{1}{\pi^3 T}N_c\sum_f\frac{|e_fH|}{2\pi}{\rm Im}\psi^{(1)}\left(\frac{1}{2}+i\frac{\beta\mu}{2\pi}\right),
\eeq
where we assumed the isospin symmetric matter, $\mu_u=\mu_d$, and $\psi^{(1)}$ is the trigamma function.
Hence the condition $\alpha_3=0$ gives $\mu=0$.
Next we evaluate $\alpha_2(H)$ in the presence of the magnetic field in 1+3 dimensions, 
\beqa
\alpha_2(H,T,\mu)&=&-2N_c\sum_{f,m}\frac{T|e_fH|}{2\pi}\int\frac{dk}{2\pi}\sum_{n} \frac{2-\delta_{n0}}{(\omega_m+i\mu)^2+k^2+2|e_fH|n}
+\frac{1}{2G}.
\eeqa
 Since it includes the quadratic divergence, we need to regularize it. 
Leaving the details in a separate paper \cite{tat2}, we plot the solution of $\alpha_2=0$ on the $T-H$ plane in the figure. 
Each point on this line is the novel tricritical point, which is also the Lifshitz point. We can also see the contribution from LLL separately. It approachs the full curve as $H$ increases, which indicates the dimensional reduction in the large $H$ limit. 

\begin{figure}
\centering
\includegraphics[width=10cm,clip]{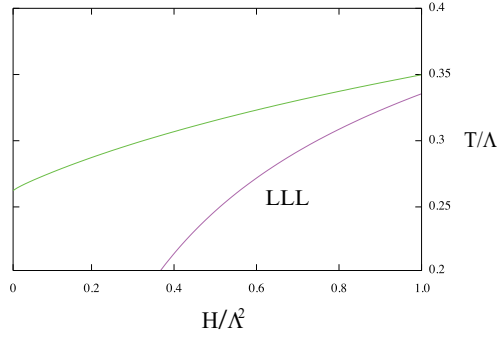}
\caption{Tricritical line on the $T-H$ plane given by the relation $\alpha_2(H,T,\mu=0)=0$. 
The proper-time regularization is used with the cut-off parameter $\Lambda$ \cite{nak}. 
The symbol "LLL" denotes the tricritical line by using only the contribution to $\alpha_2$ from LLL.}
\label{fig-1}       
\end{figure}

\section{Summary and concluding remarks}

We discussed dual chiral density wave (DCDW) in the presence of the magnetic field. 
Spectral asymmetry arises in the lowest landau level and anomalous baryon number is induced. 
Consequently DCDW phase should be greatly extended in the $\mu-T-H$ space due to the magnetic field.
It is also related to chiral anomaly \cite{son}. 

Generalized Ginzburg-Landau analysis suggests that there emerges a novel tricritical point on the $T-H$ plane. It should be interesting to explore it in the QCD lattice simulation, since it should be free from the sign problem in this plane. More elaborate study is needed to include the symmetry breaking effect.

As a phenomenological implication, we have recently suggested the rapid cooling process \cite{tat}. It should be interesting consider other consequences coming from the coupling of DCDW with the magnetic field inside compact stars or relativistic heavy-ion collisions \cite{bas2}.

%

\begin{thebibliography}{}
%
%
\bibitem{fuk}
K. Fukushima and T. Hatsuda, Rept.Prog.Phys. \textbf{74} 014001 (2011).

\bibitem{nak}
T. Tatsumi and E. Nakano, hep-ph/0408294 (2004).\\
E.Nakano and T.Tatsumi, Phys. Rev. D\textbf{71}, 114006 (2004).

\bibitem{bas} 
G. Basar and G. Dunne, Phys.Rev. D \textbf{78}, 065002 (2008).\\
G.Basar, G.V.Dunne, M.Thies, Phys. Rev. D\textbf{79}, 105012 (2009).

\bibitem{nic} 
D.Nickel, Phys. Rev. Lett. \textbf{103}, 072301 (2009); Phys. Rev. D\textbf{80}, 074025 (2009).

\bibitem{fflo}
P. Fulde and R.A. Ferrell, Phys. Rev. \textbf{135}, A550 (1964).\\
A.I. Larkin and Yu.N. Ovchinnikov, Zh.Exsp.Teor.Fiz. \textbf{47}, 1136 (1964)[Sov.Phys. JETP \textbf{20}, 762 (1965)].

\bibitem{ove}
A.W. Overhauser, Phys. Rev. \textbf{128}, 1437 (1962).

\bibitem{con}
G.J. Conduit et al., Phys. Rev. Lett. \textbf{103}, 207201 (2009).

\bibitem{alf}
M.G. Alford, J.A. Bowers, K. Rajagopal, Phys. Rev. D\textbf{63}, 074016 (2001).\\
J.A. Bowers and K. Rajagopal, Phys. Rev. D\textbf{66}, 065002 (2002).

\bibitem{lec}
For review articles, see \textit{Lecture Notes in Physics} \textbf{871} (2013).

\bibitem{bal}
G.S. Bali et al., Phys. Rev. D\textbf{86}, 071502(R) (2012).

\bibitem{nie}
A.J. Niemi and G.W Semenoff, Phys. Reports \textbf{135}, 99 (1986).

\bibitem{kle}
S.P. Klevansky and R.H. Lemmer, Phys. Rev. D\textbf{39}, 3438 (1989).

\bibitem{sug}
H. Suganuma and T. Tatsumi, Ann. Phys. \textbf{208}, 371 (1991).

\bibitem{gus}
V.P. Gusynin, V.A. Milansky and I.A. Shovkovy, Nucl. Phys. B\textbf{462}, 249 (1996).


\bibitem{sok}
A.A. Sokolov and I.M. Ternov, \textit{Radiation from Relativistic Electrons} (AIP, New York, 1986).

\bibitem{kar}
S. Karasawa and T. Tatsumi, arXiv:1307.6448 (2013).

\bibitem{fro}
I.E. Frolov et al., Phys. Rev. D \textbf{82}, 076002 (2010).

\bibitem{gol}
J. Goldstone and R.L. Jaffe, Phys. Rev. Lett. \textbf{31}, 1518 (1983).

\bibitem{mul}
P.J. Mulders, Phys. Rev. D \textbf{30}, 1073 (1984).

\bibitem{bar}
A. Barducci et al., Phys. Rev. D \textbf{51}, 3042 (1995).

\bibitem{gru}
G. Gr\"uner, \textit{Density waves in solids} (Addison-Wesley, Massachusetts, 1994).

\bibitem{sch}
V. Sch\"on and M. Thies, Phys. Rev. D \textbf{62}, 096002 (2000).

\bibitem{thi}
M. Thies, Phys. Rev. D \textbf{69}, 067703 (2004); J. Phys. A \textbf{39}, 12707 (2006). 

\bibitem{cho}
A. Chodos, I. Everding and D.A. Owen, Phys. Rev. D \textbf{42}, 2881 (1990).

\bibitem{tat2}
T. Tatsumi, K. Nishiyama and S. Karasawa, in preparation.

\bibitem{son}
D.T. Son and A.R. Zhinitsky, Phys. Rev. D \textbf{70}, 074018 (2004).\\
D.T. Son and M.A. Stephanov, Phys. Rev. D\textbf{77}, 014021 (2008).

\bibitem{tat}
T. Tatsumi and T. Muto, PoS 237 (2012).

\bibitem{bas2}
G. Basar,G.V. Dunne and D.E. Kharzeev, Phys. Rev. Lett. \textbf{104}, 232301 (2010).




\end{thebibliography}
%
%

\end{document}